\documentclass[review,10pt]{elsarticle}

\usepackage{etoolbox}
\patchcmd{\pprintMaketitle}
  {\hrule\vskip10pt}
  {\hrule\vskip10pt\ifvoid\extrainfobox\else\unvbox\extrainfobox\par\vskip10pt\fi}
  {}{}

\newsavebox\extrainfobox

\usepackage{amssymb}
\usepackage{amsbsy,graphicx,amsmath,amsfonts}
\usepackage{mathrsfs}
\usepackage{latexsym}
\usepackage{pseudocode}
\usepackage{epsfig,epsf,rotating}
\usepackage{psfrag}
\usepackage{subfigure}
\usepackage{epstopdf} 
\usepackage{float} 
\usepackage{booktabs}
\usepackage{gensymb}
\usepackage{color}
\usepackage{array}
\usepackage{arydshln}
\usepackage{txfonts}
\usepackage{siunitx}

\usepackage{bm}
\usepackage[margin=2cm]{geometry}
\biboptions{sort&compress}
\usepackage{xspace}

\makeatletter
\DeclareRobustCommand{\etc}{%
    \@ifnextchar{.}%
        {etc}%
        {etc.\@\xspace}%
}
\makeatother

\journal{Composites Part B}

\begin{document}


\begin{frontmatter}

\title{Micromechanics-based phase field fracture modelling of CNT composites}

\author[IC]{Leonel Quinteros}

\author[IC,UGR]{Enrique Garc\'{i}a-Mac\'{i}as}

\author[IC]{Emilio Mart\'{\i}nez-Pa\~neda\corref{cor1}}
\ead{e.martinez-paneda@imperial.ac.uk}

\address[IC]{Department of Civil and Environmental Engineering, Imperial College London, London SW7 2AZ, UK}
\address[UGR]{Department of Structural Mechanics and Hydraulic Engineering, University of Granada, Av. Fuentenueva sn 18002, Granada, Spain}

\cortext[cor1]{Corresponding author.}

\begin{abstract}
We present a novel micromechanics-based phase field approach to model crack initiation and propagation in carbon nanotube (CNT) based composites. The constitutive mechanical and fracture properties of the nanocomposites are first estimated by a mean-field homogenisation approach. Inhomogeneous dispersion of CNTs is accounted for by means of equivalent inclusions representing agglomerated CNTs. Detailed parametric analyses are presented to assess the effect of the main micromechanical properties upon the fracture behaviour of CNT-based composites. The second step of the proposed approach incorporates the previously estimated constitutive properties into a phase field fracture model to simulate crack initiation and growth in CNT-based composites. The modelling capabilities of the framework presented is demonstrated through three paradigmatic case studies involving mode I and mixed mode fracture conditions.
\end{abstract}



\end{frontmatter}


\section{Introduction}
\label{Introduction}

There is an increasing interest on the role of nano-modified composite materials in a number of future technologies across structural, biomedical, electronic, automotive, aircraft, and energy engineering \cite{hassan2021functional}. In particular, CNT-reinforced composite materials have attracted vast attention due to their remarkable multi-functional properties; namely, large mechanical strength, lightweight, thermochemical stability, and high electrical conductivity \cite{ESAWI20072394,Bakshi2010,samal2008carbon,yakobson2001mechanical,demczyk2002direct,ebbesen1996electrical,THOSTENSON20011899}. However, the material behaviour of CNT-based composites is not yet fully understood, and there is a need for developing models that can predict their deformation and failure.
 
The modelling of CNT-based composites is a challenging problem given their multi-scale nature and the large number of microstructural features affecting their behaviour \cite{shi2004,GARCIAMACIAS2017451}. Several approaches have been proposed to calculate the effective elastic properties of CNT-reinforced composites. Amongst them, numerical homogenisation approaches based on molecular dynamics (MD) \cite{FRANKLAND20031655,GRIEBEL20041773} or atomistic-based continuum \cite{Natsuki2004,GARCIAMACIAS2019114} are particularly popular. Nevertheless, while these techniques offer a high-fidelity representation of the microstructure, their application is often limited to the simulation of reduced sets of atoms due to computational cost. Homogenisation methods based on mean-field theory provide a computationally efficient alternative, making it possible to simulate larger microstructures in an analytical or semi-analytical manner. While relying on simplified assumptions about the interaction between the constituent phases, mean-field homogenisation has proved suitable for a wide variety of composite materials, including CNT-based composites. In this light, it is worth noting the recent work by Daghigh \textit{et al.} \cite{Daghigh2020}, who conducted non-local bending and buckling analyses of CNT-reinforced composite nanoplates resting on a Pasternak foundation using the Mori-Tanaka mean-field method. A similar approach was used by Ghasemi and co-authors \cite{Ghasemi2019} to investigate the dynamic behaviour of CNTs/fiber/polymer/metal laminated cylindrical shells. Considerable efforts have been devoted to the modelling of the specific features of CNT-based microstructures, including the variability in the geometrical properties and orientation of CNTs, filler waviness, and agglomeration. The appearance of inhomogeneous dispersions of CNTs is of particular concern since filler agglomerates act as micro-structural defects compromising the effective properties of the composite (see e.g. \cite{Koirala2021}). Filler agglomeration in clusters originates due to the large surface area of CNTs inducing strong van der Waals adhesion forces \cite{Wernik2011,ALLAOUI20021993,gkikas2015optimisation}. In the realm of mean-field homogenisation, agglomeration effects are commonly accounted for by means of the two-parameter agglomeration model proposed by Shi \textit{et al.} \cite{shi2004}. This model conceives the composite as a two-phase material, including clusters or zones of agglomerated fillers and the surrounding matrix with less dispersed fillers. The simplicity and compatibility of this approach with most mean-field homogenisation techniques have favoured its implementation in numerous research works. For instance, Moradi-Dastjerdi \textit{et al.} \cite{MoradiDastjerdi2020} investigated the electromechanical static behaviour of nanocomposite porous aggregated CNTs/polymer plates bonded between two piezoceramics faces. In their work, the elastic properties of the CNT/polymer composite were estimated by the Mori-Tanaka homogenisation model in combination with the two-parameter agglomeration model. Their results demonstrated that the flexibility of the plate is highly conditioned by the agglomeration level. Following similar homogenisation approaches, several recent contributions can be found in the literature on the analysis of agglomeration effects upon the elastic response of micro- or macroscopic structural elements, including pipes \cite{Fakhar2020}, conical panels \cite{Yousefi2020}, planar shells \cite{Daghigh2020}, or cylindrical shells \cite{Ghasemi2019,Bisheh2020}, to mention a few.

Modelling fracture poses an extra level of complexity, and the number of articles reporting on the analysis of the fracture resistance of CNT-based composites is remarkably smaller. This is largely due to the difficulties involved in the characterisation of the mechanisms governing their fracture energy properties. Most works are focused on incorporating the role of pull-out and the filler fracture mechanisms \cite{lauke1986fracture,fu1997fibre,Hsieh2011}. A pioneering work in this area is that of Mirjalili and Hubert \cite{mirjalili2010modelling}, who extended the classical formulation of bridging toughening in short-fibre composites by Fu and Lauke \cite{fu1996effects} and Kelly \cite{Kelly1970}. An analytical formulation was proposed to estimate the fracture energy of CNT-based composites including the contributions of fibre pull-out and fracture \cite{mirjalili2010modelling}. The relative contribution of these two mechanisms is governed by the so-called critical \emph{embedment length}. This length parameter governs the balance between the forces needed to break a CNT and the filler/matrix interfacial bonding force to pull it out. Below this critical length, only CNT pull-out contributes to the fracture energy of the composite, while the bridging effect is given by the sum of CNT pull-out and rupture contributions for average lengths above the critical length. The previous formulation was extended by Menna \textit{et al.} \cite{menna2016effect} to take into account statistical distributions of the length and orientation of CNTs. However, no agglomeration effects have been considered in these works. Among the few contributions reporting on such effects, it is worth noting the work by Zeidenidi \textit{et al.} \cite{ZEINEDINI201884} who experimentally investigated the fracture toughness of CNT/epoxy composites. Those authors also attempted to extend the previously mentioned formulation to account for agglomeration effects through certain empirical factors affecting the filler volume fraction and the effective Young's modulus of the composite. While the developed model was based upon strong simplifications and was dependent on experimental calibration, considerably good agreements were found with experimental data, highlighting the detrimental effect of agglomeration upon the fracture toughness of CNT-based composites. 

Fewer works have been devoted to the simulation of crack propagation in microscopic or macroscopic CNT-based composite systems and all of them involve the use of discrete computational methods. Eftekhari and Ardakani \cite{Eftekhari2014} implemented a multi-scale approach combining MD and XFEM to model the fracture behaviour of carbon CNT-reinforced concrete. Negi \textit{et al.} \cite{Negi2019} also used XFEM to investigate the fracture behaviour of thin plates doped with fully aligned CNTs, including plates with edge cracks and multiple holes. However, discrete methods are limited when dealing with complex crack topologies, arbitrary crack trajectories or the interaction of multiple cracks. Phase field fracture methods have emerged as a promising alternative to discrete approaches \cite{Bourdin2000,Wu2020,PTRSA2021}. By using an auxiliary (phase field) variable to track the interface between fractured and unbroken phases, complex cracking phenomena can be captured on the original finite element mesh. Phase field approaches are enjoying a notable popularity in fracture mechanics, spanning numerous applications; from chemo-mechanical fracture \cite{Miehe2015,CMAME2018,Wu2020b} to shape memory alloys \cite{CMAME2021,FFEMS2022}. The success of phase field fracture methods has recently been extended to composite materials (see Ref. \cite{Bui2021} for a review). Developments include the simulation of intralaminar and translaminar fracture in long-fibre composites \cite{Quintanas-Corominas2019,Zhang2021b,Nguyen2019d}, anisotropic formulations \cite{Bleyer2018,Zhang2019}, the analysis of functionally graded composites \cite{CPB2019,DT2020}, micromechanical models that explicitly resolve the microstructure \cite{Zhang2020,CST2021,CS2022} and multi-scale approaches \cite{Patil2018}, However, no phase field fracture formulation for CNT-based composites has been presented yet.

In this work, we propose the first phase field model for predicting crack nucleation and growth in CNT-based composites. The proposed approach incorporates constitutive models based on mean-field theory to predict the elastic and fracture properties of the composite material. Additionally, in view of the literature gap regarding the modelling of agglomeration effects in fracture, a novel stochastic agglomeration model is presented. The proposed agglomeration model extends the two-parameter model by Shi \textit{et al.} \cite{shi2004}, and implements a probability distribution of the number of CNTs agglomerated in bundles. To demonstrate the potential of the proposed approach, three study cases are presented. These include a single-edge notched plate subjected to uniaxial and shear loading, and a holed plate under traction. Through parametric analyses, we investigate and discuss the effects of the filler volume fraction, aspect ratio and agglomeration on the fracture behaviour of CNT-based composites.

\section{Micromechanics modelling of CNT-reinforced composites}
\label{section:2}

Here, we present our micromechanical formulation for the deformation (Section \ref{Section:2.1}) and fracture (Section \ref{Section:2.2}) behaviours of the composite material.

\subsection{Effective elastic properties}
\label{Section:2.1}
\subsubsection{Effective elastic moduli using a double-inclusion model}
\label{Section:2.1-double_inclusion}

\begin{figure}[h]
\centering
\includegraphics[scale=0.85]{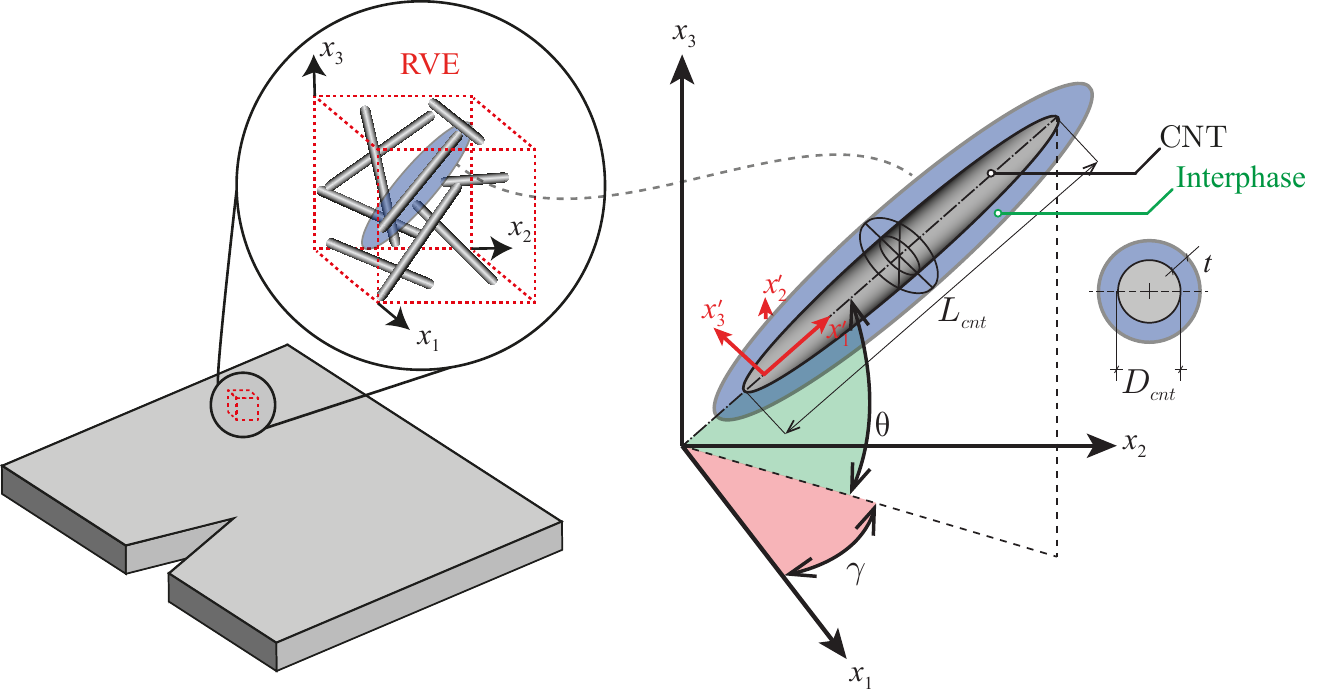}
\caption{Schematic representation of a CNT-based composite plate, a representative volume element (RVE), and the local orientation of a CNT defined by the Euler angles, $\theta$ and $\gamma$.}
\label{fig:fig1}
\end{figure}

To estimate the elastic properties of CNT-based composites, let us consider the representative volume element (RVE) shown in Fig. \ref{fig:fig1}. This comprises the matrix phase, randomly oriented CNTs, and the interface between them, which are denoted by indexes $m$, $p$, and $i$, respectively. This RVE is assumed to contain a sufficient number of fillers such that the overall properties of the composite are statistically represented. A local coordinate system K$'\equiv\left\{0;x'_1x'_2x'_3\right\}$ is fixed at each particle, and two Euler angles, $\theta$ and $\gamma$, are defined to describe the relative orientation of the fillers with respect to the global coordinate system. The geometrical dimensions of the CNTs are assumed constant throughout the RVE, including their length $L_{cnt}$, diameter $D_{cnt}$, and the interphase thickness $t$.
  Following the notation of Hori and Nemat-Nasser \cite{Hori1993}, a CNT and its surrounding interphase can be modelled as a double inclusion as shown in the right part of Fig. \ref{fig:fig1}. Denoting the linear elastic tensors of the constituent phases as $\mathbf{C}_m$, $\mathbf{C}_p$, and $\mathbf{C}_i$, and the corresponding volume fractions as $f_m$, $f_p$ and $f_i$, the effective constitutive tensor of the three-phase composite can be obtained as \cite{Xu2017a,GARCIAMACIAS201849}:
\begin{equation}
   \bar{\mathbf{C}}=(f_{m}\mathbf{C}_{m}+ f_{i}\langle\mathbf{C}_i\colon \mathbf{A}_i\rangle+f_p \langle \mathbf{C}_{p}\colon \mathbf{A}_p\rangle)\colon (f_m \mathbf{I}+f_i \langle\mathbf{A}_i\rangle +f_p \langle\mathbf{A}_p\rangle)^{-1}
    \label{eq:rule_of_mix_nd},
\end{equation}

\noindent where $\mathbf{I}$ is the fourth-order identity tensor, whereas $\mathbf{A}_i$ and $\mathbf{A}_p$ refer to the concentration tensors for the interphases and the inclusions, respectively. These quantities can be written as a function of the dilute concentration tensors $\mathbf{A}_i^{dil}$ and $\mathbf{A}_p^{dil}$ as \cite{Hori1993}:
\begin{equation}
    \mathbf{A}_{\alpha}=\mathbf{A}_{\alpha}^{dil} \colon (f_{m} \mathbf{I}+f_{i}\mathbf{A}_{i}^{dil}+f_{p}\mathbf{A}_{p}^{dil})^{-1}, \quad \alpha=p,i
    \label{eq:A_chi}
\end{equation}
where
\begin{equation}
\mathbf{A}_{\alpha}^{dil} = \mathbf{I}+\mathbf{S}\colon \mathbf{T}_{\alpha}, \quad \alpha=p,i
\label{eq:A_chi_dil}
\end{equation}
\begin{equation}
    \mathbf{T}_{\alpha}=-(\mathbf{S}+\mathbf{M}_{\alpha})^{-1},\:\:\: \alpha=p,i
    \label{eq:T_chi}
\end{equation}
\begin{equation}
    \mathbf{M}_{\alpha}=(\mathbf{C}_{\alpha}-\mathbf{C}_{m})^{-1}\colon \mathbf{C}_{m}, \quad \alpha=p,i
    \label{eq:M_chi}
\end{equation}

\noindent Here, $\mathbf{S}$ corresponds to Eshelby’s tensor for an spheroidal particle, determined by the aspect ratio of the CNTs, $\kappa=L_{cnt} / D_{cnt}$, and by the Poisson's ratio of the matrix, $\nu_{m}$. The reader is referred to Ref. \cite{Mura1987} for explicit definitions of $\mathbf{S}$ for a variety of filler geometries. The angle brackets $\langle \cdot \rangle$  in Eq. (\ref{eq:rule_of_mix_nd}) denote the orientational average over the entire space of Euler angles weighted by an orientation distribution function (ODF), $\Omega(\gamma,\theta)$. Specifically, the orientational average of a certain function $F(\gamma,\theta)$ reads:
\begin{equation}
\left\langle F \right\rangle=\int_0^{2\pi}\int_0^{\pi/2}F(\gamma,\theta)\Omega(\gamma,\theta)\sin(\theta) \, \textrm{d}\theta\textrm{d}\gamma.
\end{equation}

In the particular case of random filler orientations, any orientation is equally probable and the ODF takes the shape of a uniform probability distribution with a constant value $\Omega(\gamma,\theta)=1/2\pi$.

It remains to compute the volume fraction of the interphases $f_{i}$. Assuming penetrable soft interphases, $f_{i}$ can be calculated using the formulation derived by Xu \textit{et al.} \cite{Xu2016a} as:
 \begin{equation}
\label{Eq:softvi}
f_i=(1-f_{p})\left(1-\exp \left\{-\frac{6f_p}{1-f_{p}}\left[\frac{\eta}{n(\kappa)}+\left(2+\frac{3f_{p}}{n^2(\kappa)(1-f_{p})}\right)\eta^2+ \frac{4}{3}\left(1+\frac{3f_{p}}{n(\kappa)(1-f_{p})}\right)\eta^3\right]\right\}\right),
\end{equation}
 
\noindent with $\eta$ being the ratio of the interfacial thickness $t$ and the equivalent diameter $D_{eq}$ (i.e.~$\eta=t/D_{eq}$). The equivalent diameter denotes the diameter of an equivalent sphere with the same volume as the particles \cite{Beddow2018}. In the case of CNTs with aspect ratio $k>1$, $D_{eq}$ can be determined as $D_{eq}=D_{cnt}\kappa^{1/3}$. Finally, the term $n(\kappa)$, denotes the sphericity of the CNTs and is defined as the ratio between the surface areas of the equivalent spheres and the particles, which reads \cite{WANG2021112862}:
\begin{equation}\label{hardvi2}
n(\kappa)=\frac{2\kappa^{2/3}\tan \varphi}{\tan \varphi+\kappa^{2}\varphi},
\end{equation}

\noindent with $\varphi=\textrm{arcos} (\beta)$, and $\beta=1/\kappa$. 
 
\subsubsection{Agglomeration of CNTs}
\label{2.1.1:Agglomeration of CNTs_young}

In order to incorporate agglomeration effects into a mean-field homogenisation framework, the two-parameter agglomeration model by Shi \textit{et al.} \cite{shi2004} is implemented. This approach conceives the composite as a two-phase material, including bundles with high filler contents and the lightly loaded surrounding matrix. Therefore, the volume of the RVE, $V$, and the volume of CNTs $V_r$ can be written as:
\begin{equation}
    V=V_{bundles}+V_{matrix},\,\,\,\,\,\,\,\,\,\,\,\, V_r=V_{r}^{bundles}+V_{r}^{m},
\end{equation}

\noindent where $V_{bundles}$ and $V_{matrix}$ refer to the volume of the bundles and the matrix, respectively, whereas  $V_{r}^{bundles}$ and $V_{r}^{matrix}$ stand for the CNT concentrations in the bundles and in the matrix, respectively. Now, two agglomeration parameters, $\chi$ and $\zeta$, can be defined as follows:
\begin{equation}
    \chi=\frac{V_{bundles}}{V},\,\,\,\,\,\,\,\,\,\,\,\, \zeta=\frac{V_{r}^{bundles}}{V_{r}},
    \label{Eq:chi_and_zeta}
\end{equation}

\noindent and, considering the total CNTs volume fraction $f_{p}=V_r/V$,  the filler volume fractions in the bundles $f_{bundle}$ and in the matrix $f_{matrix}$ can be obtained as:
\begin{equation}
    f_{bundles}=\frac{V_{r}^{bundles}}{V_{bundles}}=\frac{\zeta}{\chi}f_{p},\,\,\,\,\,\,\,\,\,\,\,\, f_{matrix}=\frac{V_{r}^{matrix}}{V_{matrix}}=\frac{1-\zeta}{1-\chi}f_{p}.
    \label{Eq: f_bundles_matrix}
\end{equation}

\begin{figure}[H]
\centering
\includegraphics[scale=0.85]{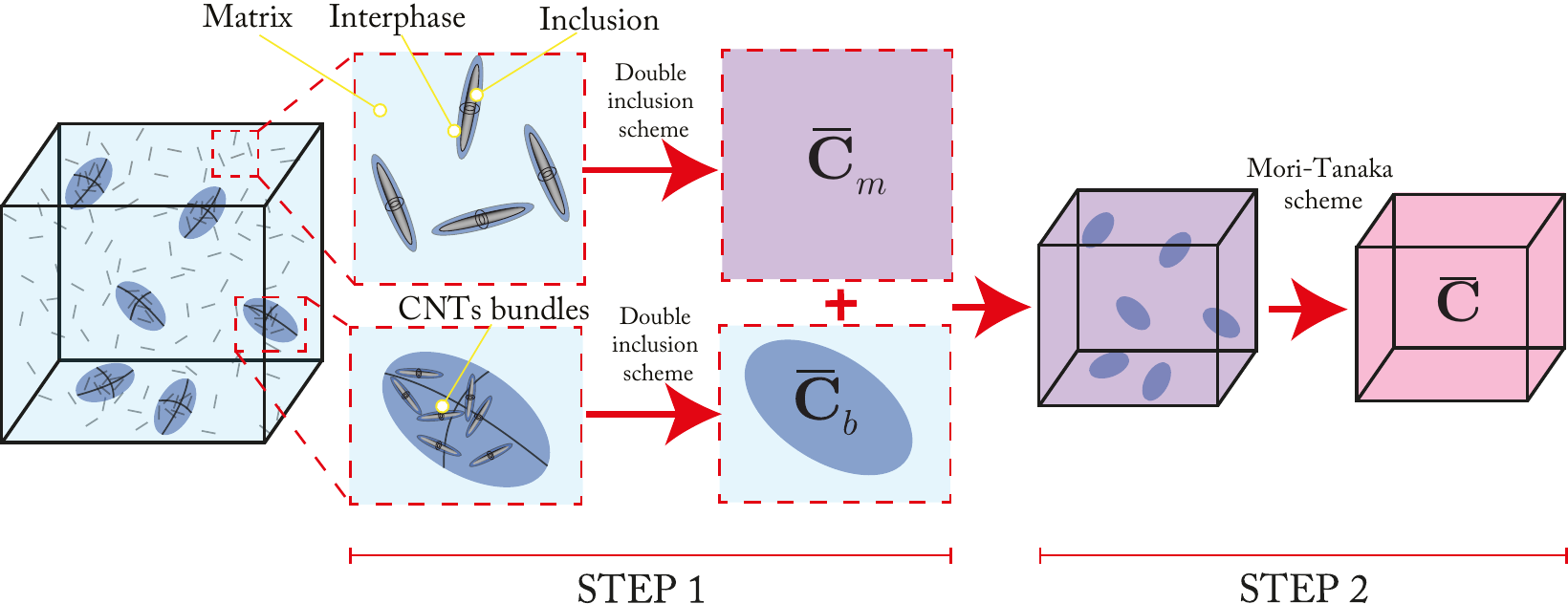}
\caption{Scheme of the two-step micromechanics model to incorporate agglomeration effects on the elastic properties of CNT-based composites.}
\label{fig:agglomeration_model}
\end{figure}

On this basis, a two-step homogenisation model inspired by Ref. \cite{GARCIAMACIAS201849} is implemented as sketched in Fig. \ref{fig:agglomeration_model}. The first step consists in the estimation of the effective properties of the clusters and the homogeneously dispersed CNTs in the surrounding matrix, separately. To do so, the double-inclusion model previously overviewed in Section \ref{Section:2.1-double_inclusion} (Eq. (\ref{eq:rule_of_mix_nd})) is applied to the bundles and the surrounding matrix independently, with the only difference being the volume fraction of the reinforcing fillers. Specifically, values of $f_p=f_{bundles}$ and $f_p=f_{matrix}$ are considered for the bundles and the surrounding matrix, respectively. The resulting constitutive tensors of the bundles and the surrounding matrix are denoted as $\bar{\mathbf{C}}_{b}$ and $\bar{\mathbf{C}}_{m}$, respectively. Next, the overall constitutive tensor of the composite is obtained in the second step considering bundles as spherical inclusions and the lightly loaded matrix as the matrix phase. In this case, the volume fraction of the bundles is given the by parameter $\chi$, as previously reported in Eq. (\ref{Eq:chi_and_zeta}). Considering the constitutive tensors obtained in the first step, $\bar{\mathbf{C}}_{b}$ and $\bar{\mathbf{C}}_{m}$, the overall constitutive tensor can be estimated by applying the Eshelby-Mori-Tanaka model as \cite{Mori1973}:
\begin{equation}\label{Agglom2param}
    \bar{\mathbf{C}}=\bar{\mathbf{C}}_{m}+\chi(\bar{\mathbf{C}}_{b}-\bar{\mathbf{C}}_{m}) \colon \mathbf{A},
\end{equation}

\noindent where,
\begin{equation}
\mathbf{A}=\mathbf{A}_{dil}\colon \left[ (1-\chi)\mathbf{I}+\chi \mathbf{A}_{dil}  \right]^{-1},
\end{equation}

\noindent and
\begin{equation}
    \mathbf{A}_{dil}=(\mathbf{I}+\mathbf{S}_b\colon \mathbf{C}_{m}^{-1})\colon (\mathbf{C}_{b}-\mathbf{C}_{m})^{-1}.
    \label{A_dil_mori}
\end{equation}

Eshelby's tensor $\mathbf{S}_b$ in Eq. (\ref{A_dil_mori}) depends on the geometry of the clusters, which is assumed to be spheroidal. Nonetheless, the formulation in Eq. (\ref{Agglom2param}) is general, and different geometries of clusters can be accounted for by implementing the corresponding Eshelby's tensor.

\subsection{Fracture energy formulation}

\begin{figure}[h]
\centering
\includegraphics[scale=0.95]{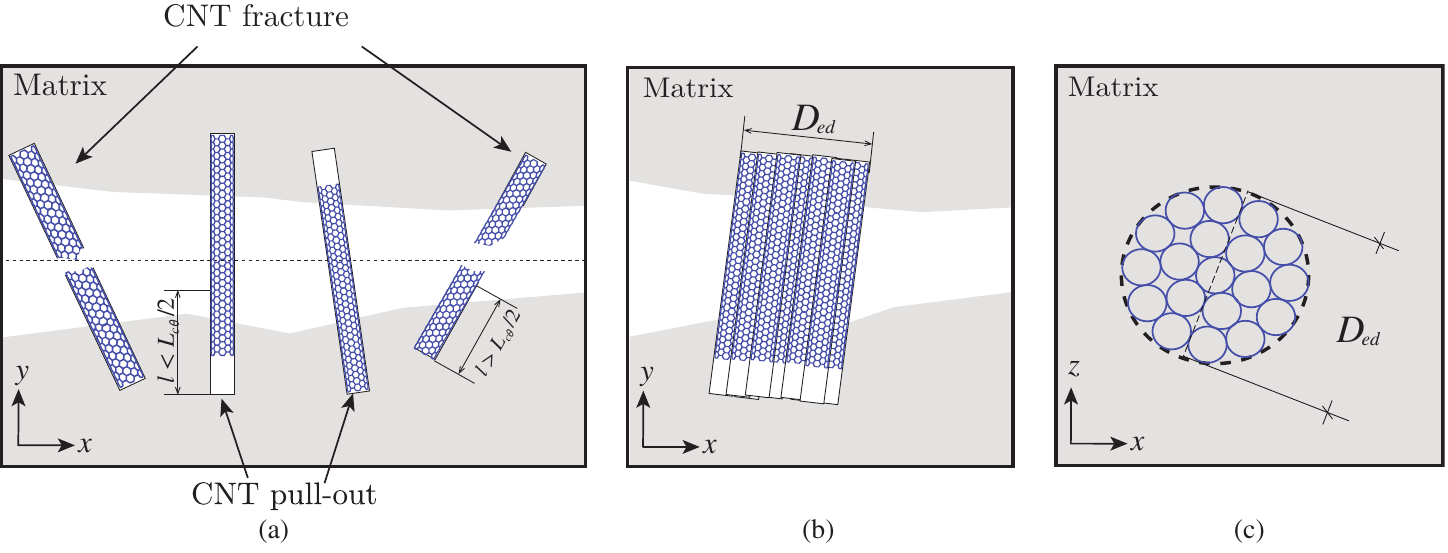}
\caption{Micromechanics of CNT composite fracture: (a) Schematic representation of the CNT-induced bridging mechanisms (pull-out and CNT rupture), (b) agglomeration of CNTs, and (c) cross-section plane of the agglomeration model (involving 19 CNTs).}
\label{fig:crack_mechanism_agg}
\end{figure}

As outlined in Fig. \ref{fig:crack_mechanism_agg} two main elements set the basis for our micromechanical homogenisation of the fracture behaviour: (i) a toughening contribution from pull-out and CNT rupture mechanisms, and (ii) the role of CNT agglomeration.

\label{Section:2.2}
\subsubsection{Fracture Energy: pull-out and fracture}

The fracture resistance of CNT-based composites is mainly governed by two contributions: the toughness of the matrix, $G_0$, and the fibre bridging toughening mechanisms; the latter are accounted for here through the term $G_{PF}$. Accordingly, the CNT-based composite toughness equals,
\begin{equation}
    G_{c}=G_{0}+G_{PF},
\end{equation}

\noindent where $G_{PF}$ incorporates the two main fibre toughening mechanisms identified in the literature \cite{menna2016effect,fu1996effects}: the fracture and pull-out of CNTs (see Fig. \ref{fig:crack_mechanism_agg}a). Their relative contribution can be quantified by means of the critical length $L_{c\theta}$ \cite{li1991micromechanical}, which can be defined from applying a simple force balance to a CNT as:
\begin{equation}\label{Lcrit}
    A_{cnt} \sigma_{ult\theta}=L_{c\theta} P_{cnt} \tau_{int}  e^{\mu \theta},
\end{equation}

\noindent where $A_{cnt}$ is the area of the CNT cross-section, $P_{cnt}$ is the perimeter of the cross-section, $\tau_{int}$ is the interfacial frictional shear stress, and $\sigma_{ult \theta}$ is the fracture stress of oblique fibres. For brittle fibres, $\sigma_{ult \theta}$ is given by \cite{PIGGOTT1974457}:
\begin{equation}
    \sigma_{ult\theta} = \sigma_{ult}\left[1-A \tan(\theta)\right],
    \label{Eq:force_balance}
\end{equation}

\noindent where $\sigma_{ult}$ is the fracture strength of a CNT, and $A$ is a constant determining the fibre inclined strength. Then, if the embedment length $l$ of a CNT is lower than the critical length, i.e.~$l<L_{c\theta}/2$, the CNT will pull out. Conversely, the CNT rupture mechanism will take place if $l \geq L_{c\theta}/2$. Thus, the work of fracture of a single CNT can be defined in a piecewise fashion as \cite{menna2016effect}:
\begin{equation}
W(l,\theta) = \begin{cases}l^{2}\tau_{int}P_{cnt} \exp(\mu \theta)/2 &\mbox{if }l<\frac{L_{c\theta}}{2} \\
 A_{cnt} \sigma_{ult}^{2} L_{cnt}/\left(2E_{cnt}\right) & \mbox{if } l\geq\frac{L_{c\theta}}{2} \end{cases},
 \label{Eq:W_en}
\end{equation}

\noindent where $E_{cnt}$ is the Young's modulus of the CNT, and $\mu$ is the snubbing friction coefficient for misaligned CNTs \cite{li1991micromechanical}. Assuming CNTs present a cylindrical geometry, $A_{cnt}=\pi D_{cnt}^{2} /4$ and $P_{cnt}=\pi D_{cnt}$. Then, the fracture energy considering straight CNTs can be obtained as \cite{fu1997fibre}:
\begin{equation}
    G_{PF}= \frac{2f_{p}}{A_{cnt}L_{cnt}}\int_{\theta=0}^{\pi/2} \int_{l=0}^{L_{cnt}/2} \cos(\theta) W(l,\theta)g(\theta) \, \rm{d}l \rm{d}\theta
    \label{Eq:G_c},
\end{equation}

\noindent where $g(\theta)$ is an orientation distribution function accounting for the orientation of CNTs. Although the orientation of CNTs is three-dimensional in nature, one single angle $\theta$ between the loading direction and the fibre axis has been reported to suffice to capture the effect of the fibre orientation in short-fibre composites \cite{fu1996effects,jain1992effect}. In statistical terms, $g(\theta)$ can be defined to describe the planar orientation distribution of the CNTs as \cite{advani1987use,xia1995flexural}:
\begin{equation}
    g(\theta)=\frac{\sin(\theta)^{2p-1}\cos(\theta)^{2q-1}}{\int_{\theta_{min}}^{\theta_{max}} \left[\sin (\theta)^{2p-1}\cos(\theta)^{2q-1}\right]\text{d}\theta}.
    \label{Eq:orientation_pdf}
\end{equation}

\noindent The orientation angle $\theta$ in Eq. (\ref{Eq:orientation_pdf}) ranges between $\theta_{min}$ and $\theta_{max}$, which are the minimum and maximum CNT inclinations with respect to the load direction, while $p\geq 1/2$ and $q\geq 1/2$ are parameters that determine the shape of the PDF $g(\theta)$ \cite{fu1996effects}.

\subsubsection{CNT agglomeration effects}
\label{Sect:agglomeration_fracture}

The previous formulation assumes well-dispersed CNTs. However, as discussed in Section \ref{2.1.1:Agglomeration of CNTs_young}, CNTs tend to agglomerate in bundles with the subsequent detrimental effect on the fracture toughness \cite{gojny2005influence,gojny2004carbon}. In order to include such effects, a new agglomeration formulation is proposed in this work. Since no assumptions about the filler shape have been made in Eqs. \eqref{Eq:W_en} and \eqref{Eq:G_c}, agglomeration effects are incorporated by considering equivalent fibers with cross-sections corresponding to the sum of the sections of the CNTs forming bundles. Specifically, we assume that the bundle cross-section is constituted of smaller CNTs of diameter $D_{cnt}$ packed in a bigger circle of diameter $D_{ed}$, as sketched in Fig. \ref{fig:crack_mechanism_agg}(b), following the so-called ``equal circles packed in circle problem'' \cite{GRAHAM1998139} pattern depicted in Fig. \ref{fig:crack_mechanism_agg}(c). The perimeter of the CNT bundle is approximated to the perimeter of the enclosing circle with diameter $D_{ed}$, while the bundle cross-section area is the area of a single CNT multiplied by the number of CNTs forming the agglomerate ($N$). Some example values of the equal circles packed in circle problem are shown in Table \ref{Tab:ex_values}, the ratio is defined as $R=D_{ed}/D_{cnt}$, while the density is expressed as $ \rho_{A}= (NA_{cnt})/A_{ed}$, where $A_{ed}$ is the enclosing circle area.


\begin{table}[h]								
\newcommand\Tstrut{\rule{0pt}{0.3cm}}         
\newcommand\Bstrut{\rule[-0.15cm]{0pt}{0pt}}   
 \footnotesize												
  \caption{Ratio and density in function of the number of CNTs inside the enclosing circle diameter}												
 \vspace{0.1cm}												
 \centering												
 \begin{tabular}{llllllll} 												
	\hline 	
\textbf{Number of CNTs} $N$ &1&2&5&10&20&50&100\\
\textbf{Ratio}   $R=D_{ed}/D_{cnt}$ &1&2&2.701&3.813&5.122&7.947&11.082\\
\textbf{Density} $\rho_{A}=\left(N A_{cnt}\right) /A_{ed}$ & 1&0.5&0.685&0.687&0.762&0.791&0.814\Bstrut\\ 
\hline 												
 \end{tabular}												
\label{Tab:ex_values}												
 \end{table}
 
The perimeter of the agglomerates is defined as $P_{agg}(N,D_{cnt})=\pi(D_{ed})=\pi(RD_{cnt})$, while their area is a function of the number CNTs forming the agglomerates $A_{agg}(N,D_{cnt})=NA_{cnt}$. Now, with functions $P_{agg}(N,D_{cnt})$ and $A_{agg}(N,A_{cnt})$ defined from the data in Table \ref{Tab:ex_values} (see Ref. \cite{GRAHAM1998139}), the fracture energy for agglomerated CNTs can be estimated as:
\begin{equation}
    G_{PFN}^{agg}(N)= \frac{2f_{p}}{A_{agg}(N,A_{cnt})L_{cnt}}\int_{\theta=0}^{\pi/2} \int_{l=0}^{L_{cnt}/2} \cos(\theta) W(l,\theta,N)g(\theta)\rm{d}l \rm{d}\theta
    \label{Eq:G_c_agg},
\end{equation}

\noindent with $W(l,\theta,N)$ defined as:
\begin{equation}
W(l,\theta,N) = \begin{cases}l^{2}\tau_{int}P_{agg}(N,D_{cnt}) \exp(\mu \theta)/2 &\mbox{if }l<\frac{L_{c\theta}}{2} \\
 A_{agg}(N,A_{cnt}) \sigma_{ult}^{2} L_{cnt}/\left(2E_{cnt}\right) & \mbox{if } l\geq\frac{L_{c\theta}}{2} \end{cases}.
 \label{Eq:W_agg}
\end{equation}

Note that, in the case of a fracture energy dominated by CNT fracture ($l\geq L_{c\theta}/2$), the area term $A_{agg}$ drops out from Eq. (\ref{Eq:G_c_agg}) when including Eq. (\ref{Eq:W_agg}). This is due to the circumstance that the fracture energy attained by CNTs forming aggregates equals the work done to fracture the CNTs in case that they were not agglomerated. I.e., in the context of fracture, the role of agglomeration is to facilitate CNT pull-out. 
In order to simulate the randomness in the filler agglomeration, the number $N$ of CNTs clustered forming the aggregates is defined in statistical terms through a Weibull PDF $p(D)$ with shape parameters $\lambda$ and $k$. Shape parameters $\lambda$ and $k$ can be calculated numerically given the mean $N_{\mu}$ and the standard deviation $N_{\sigma}$ of the statistical distribution of $N$. On this basis, the fracture energy contributed by the filler aggregates can be estimated by integrating $G_{PFN}^{agg}$ between the minimum and maximum possible number of CNTs forming bundles, $N_{min}$ and $N_{max}$, and weighted by $p(N)$ as:
\begin{equation}\label{contrib2}
    G_{PF}^{agg}=\int_{N_{min}}^{N_{max}}G_{PFN}^{agg}p(N)\rm{d}N.
\end{equation}

Finally, in order to combine the fracture energy contributions by agglomerated and non-agglomerated fillers, the two-parameter agglomeration model previously presented in Section \ref{2.1.1:Agglomeration of CNTs_young} is also introduced here. To do so, the classical rule of mixtures is used to combine the fracture energies from Eqs. (\ref{Eq:G_c}) and (\ref{contrib2}), leading to:
\begin{equation}
    G_{c}=G_{0}+(1-\zeta)G_{PF}(f_{p})+\zeta G_{PF}^{agg}(f_{p}),
    \label{GCbundle}
\end{equation}

Note that the formulation in Eq. \eqref{GCbundle} only depends on $\zeta$, which is due to the linear relation between the volume fraction $f_{p}$ and the fracture energy $G_{PFN}^{agg}$ in Eq. \eqref{Eq:G_c_agg}.

\subsubsection{Experimental verification}

\begin{figure}[h]
\centering
\includegraphics[scale=1.0]{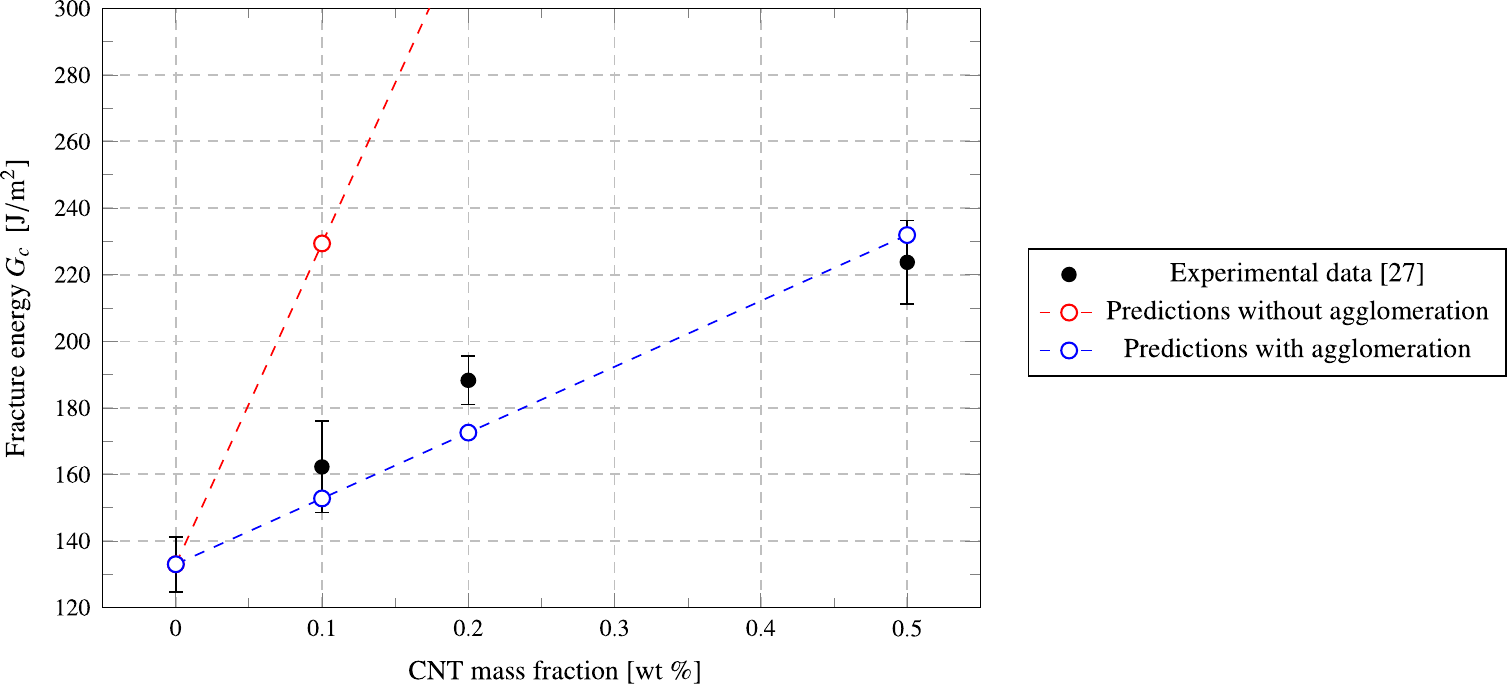}
\caption{Validation of the fracture energy formulation with and without agglomeration effects presented in Sect. \ref{Sect:agglomeration_fracture}, using the experimental fracture energy versus CNT mass fraction data provided by Hsieh \textit{et al.} \cite{Hsieh2011}. The modelling results are obtained using Eqs. (\ref{Eq:G_c}) and (\ref{Eq:G_c_agg}) for the predictions without and with agglomeration effects, respectively.}
\label{Fig:experimental_val}
\end{figure}

We proceed to validate our new formulation for incorporating agglomeration effects into the fracture energy of CNT-based composites. To achieve this, the experimental data provided by Hsieh \textit{et al.} \cite{Hsieh2011} is used. Specifically, they measured the fracture energy sensitivity to the CNT mass fraction (in wt \%) using single-edge notch-bend samples containing a sharp crack. The materials employed are an anhydride-cured epoxy polymer and multi-walled CNTs with a length of 120 $\mu$m and a diameter of 120 nm. Our choice of model parameters builds upon the data provided by Hsieh \textit{et al.} \cite{Hsieh2011}, including a CNT strength of 35 GPa, an interfacial shear strength of 47 MPa, an orientation limit angle $A=0.083$, and densities of the CNTs and the epoxy of 1.8 g/ml and 1.2 g/ml, respectively. The orientation parameters read $p=20.5$ and $q=0.5$, and the agglomeration behaviour is described by $\zeta=0.9$, $N_{min}=1$, $N_{max} = 99$, $N_{\mu}=91$, and $N_{\sigma}=2$. The predictions of our model with and without considering the effect of CNTs agglomeration are shown in Fig. \ref{Fig:experimental_val}, together with the experimental data by Hsieh \textit{et al.} \cite{Hsieh2011}. It can be shown that taking into consideration agglomeration effects is key to capture the experimental data, and that our agglomeration-enhanced micromechanical model delivers a good agreement with experiments.

\section{A phase field fracture formulation for CNT-based composites}
\label{Section:3}

Consider a solid domain $\Omega$ which includes a discontinuous surface $\Gamma$. To characterise a discrete crack, an auxiliary phase field variable $\phi$ is defined taking values from $\phi=0$ to $\phi=1$, which correspond to the intact and fully broken states of the material, respectively. The phase field provides a regularisation of the crack surface, whose size is governed by the length scale $\ell$. Accordingly, the fracture energy of the solid is approximated as \cite{Bourdin2000,PTRSA2021}:
\begin{equation}
    \int_{\Gamma} G_{c} \text{d} \Gamma \approx \int_{\Omega}G_{
    c}\Gamma_{\ell}(\ell,\phi)\text{d}\Omega = \int_{\Omega} G_{c}\left( \frac{1}{2\ell} \phi^{2}+\frac{\ell}{2} \lvert \nabla \phi \rvert ^{2}\right)\text{d}\Omega,
\end{equation}

\noindent Then, the total potential energy of the solid reads,
\begin{equation}
    \Psi=\int_{\Omega}\left( \left( 1 - \phi \right)^2 \psi+G_{c}\left(\frac{1}{2 \ell} \phi^{2}+\frac{\ell}{2}|\nabla \phi|^{2}\right)\right) \mathrm{d} \Omega,
    \label{eq2}
\end{equation}

\noindent where $\psi$ is the strain energy density of the solid. Both the fracture driving force ($\psi$) and the fracture resistance ($G_c$) are dependent on the underlying CNT distribution, as described in Section \ref{section:2}. The strong form of the coupled deformation-fracture system can be readily obtained by applying Gauss theorem to \eqref{eq2}, rendering:
\begin{equation} \label{eq:StrongForm}
    \begin{aligned}
\nabla \cdot \left[ \left( 1 - \phi \right)^2 \bm{\sigma} \right] &=0 \quad \text { in } \quad \Omega\\
G_{c}\left(\frac{1}{\ell} \phi-\ell \nabla^2 \phi\right)-2(1-\phi) \psi &=0 \quad \text { in } \quad \Omega \\
\end{aligned}
\end{equation}

\noindent where $\bm{\sigma}$ denotes the Cauchy stress tensor. The system (\ref{eq:StrongForm}) is solved in a monolithic manner, using a quasi-Newton method \cite{Wu2020a,TAFM2020}. Also, a history field is defined to enforce damage irreversibility \cite{Miehe2010a}. 

\section{Results and discussion}
\label{Section:5-Results}

The modelling capabilities of the framework presented in Sections \ref{section:2} and \ref{Section:3} is demonstrated by simulating the fracture behaviour of epoxy doped with multiwalled carbon nanotubes (MWCNTs). The material parameters, which are kept constant throughout the study, are reported in Table \ref{fittedparam}. For better illustration, this section is divided into two parts. Firstly, the effective properties of CNT-reinforced composites computed by the micromechanics method from Section \ref{section:2} are discussed in Section \ref{Section:5.1-Effective_properties}. Secondly, Section \ref{Section:5.1-Benchmarks} reports the macroscopic crack propagation analysis of three representative case studies.

\begin{table}[h]								
\newcommand\Tstrut{\rule{0pt}{0.3cm}}         
\newcommand\Bstrut{\rule[-0.15cm]{0pt}{0pt}}   
 \footnotesize												
  \caption{Micromechanical variables of MWCNT/epoxy composites. Taken from Refs. \cite{menna2016effect} and \cite{garcia2018mwcnt}.}										
 \vspace{0.1cm}												
 \centering												
 \begin{tabular}{lll} 												
	\hline 	
Name & Symbol & Value \Tstrut\\
  \hline
Length of CNTs          & $L_{cnt}$	&	3.21	$\muup$m	\\ 
Outer diameter of CNTs & $D_{cnt}$	&	10.35	nm			\\
CNT volume fraction&	$f_{cnt}$	&	1 \%	 \\
Elastic modulus of CNTs&	$E_{cnt}$	&	700	GPa	 \\
Elastic modulus of epoxy&$E_m$	&	2.5	GPa	\\
Possion's ratio	of CNTs   & $\nu_{cnt}$	&	0.3\\
Possion's ratio of epoxy & $\nu_m$	&	0.28		\\
Interphase thickness & $t$	&	31.00	nm \\
Elastic modulus of interphases &	$E_i$	&	2.17	 GPa	\\
Strength of CNTs & $\sigma_{cnt}$ & 35 GPa \\
Agglomeration parameter $\chi$ & $\chi$ & 0.2\\
Agglomeration parameter $\zeta$ & $\zeta$ & 0.4\\
Interfacial shear strength & $\tau_{cnt}$ & 47 MPa \\
Experimental orientation limit angle & $A$ & 0.083\\
Fracture energy of pristine epoxy & $G_{0}$ & 133 J/m$^2$ \\
Mean value of the PDF of the number of CNT forming agglomerates &$N_{\mu} $& $10$ \\
Standard deviation of the PDF of the number of CNT forming agglomerates &$N_{\sigma} $& $0.1N_{\mu}$ \\
Minimum number of CNTs forming agglomerates & $N_{Min} $& $1$ \\
Maximum number of CNTs iforming agglomerates & $N_{Max} $& $50$ \\
Minimum CNT orientation angle & $\theta_{min}$ & 0 \\
Maximum CNT orientation angle & $\theta_{max}$ & $\pi/2$ 
 \Bstrut\\
\hline 												
 \end{tabular}												
\label{fittedparam}												
 \end{table}
 
\subsection{Effective mechanical properties of CNT-reinforced composites}
\label{Section:5.1-Effective_properties}

\subsubsection{Uniformly dispersed CNTs}

\begin{figure}[h!]
\centering
\includegraphics[scale=1.0]{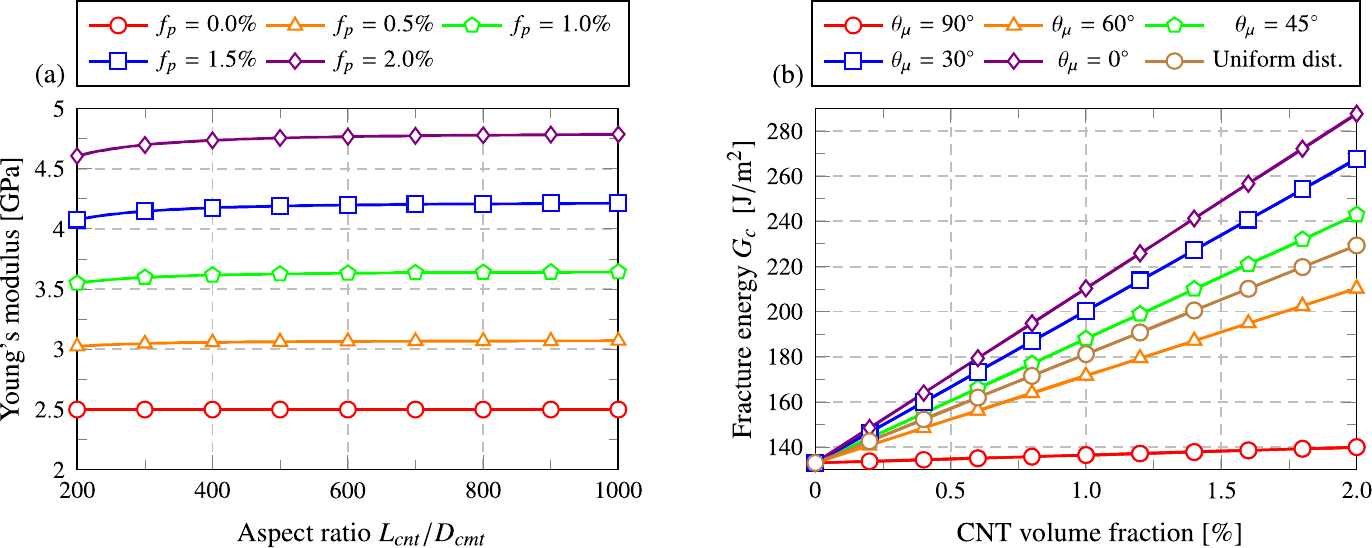}
\caption{Effective elastic modulus of MWCNT/epoxy composites as a function of the CNT aspect ratio $L_{cnt}/D_{cnt}$ for different filler volume fractions and assuming uniformly dispersed CNTs (a). Fracture energy of MWCNT/epoxy composites as a function of the filler aspect ratio $L_{cnt}/D_{cnt}$, considering different CNT contents (b).} 
\label{Fig:WDYoungAR}
\end{figure}

First, we study the mechanical properties of composites doped with uniformly dispersed CNTs. Fig. \ref{Fig:WDYoungAR}a depicts the effective Young's modulus as a function of the filler aspect ratio $L_{cnt}/D_{cnt}$, for different CNT volume fractions. The Young's modulus exhibits a slight increase at low filler aspect ratios although it tends to converge to a stable value for moderate to large aspect ratios ($L_{cnt}/D_{cnt} \geq 400$). A high sensitivity to the CNT volume content is observed. For instance, it is found that the addition of only a 0.5\% volume fraction of CNTs leads to an increase of around 20\% relative to the elastic modulus of pristine epoxy.

\begin{figure}[h]
\centering
\includegraphics[scale=1.0]{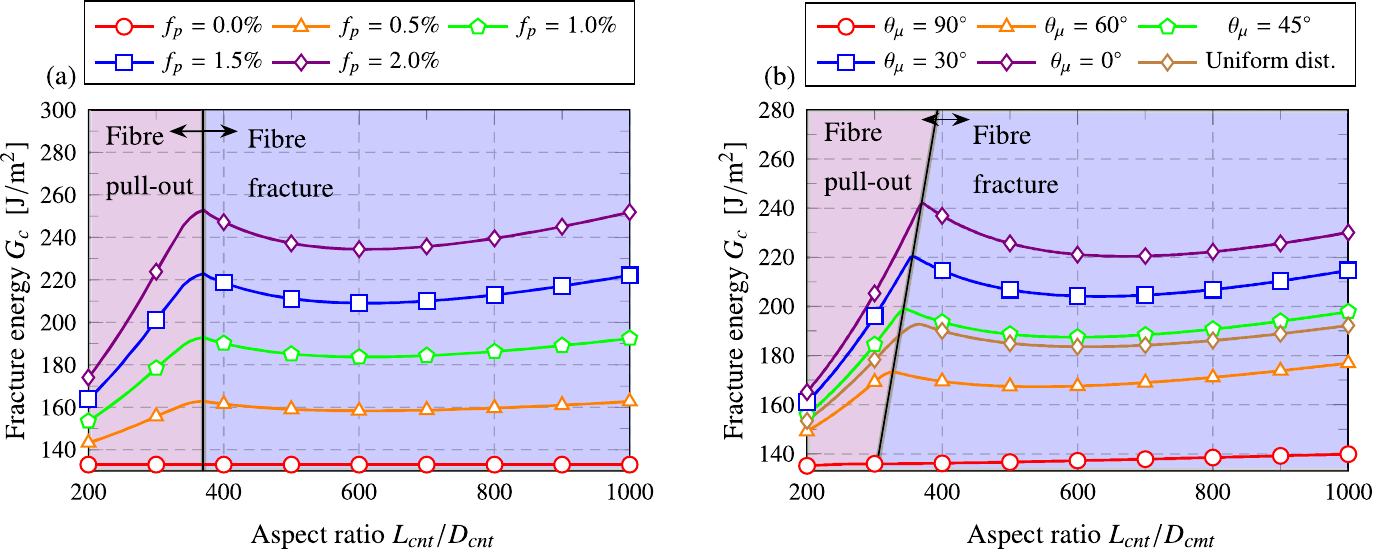}
\caption{Fracture energy of MWCNT/epoxy composites as a function of the filler aspect ratio $L_{cnt}/D_{cnt}$, considering different CNT filler contents (a), and different CNT orientation distributions (b).}
\label{Fig:WDfracture}
\end{figure}

The fracture energy exhibits a more complex behaviour. Firstly, the effect of filler orientation upon the fracture energy is investigated in Fig. \ref{Fig:WDYoungAR}b. Results are shown for selected values of the CNT distribution mean angle direction $\theta_{\mu}$. The parameters $p$ and $q$ are obtained for a fixed standard deviation $\theta_{\sigma}=0.05\frac{\pi}{2}$ and mean $\theta_{\mu}$ values ranging between $0^{\circ}$ and $90^{\circ}$. In addition, the case of a perfectly random distribution is included by fixing $p=q=1/2$. It is observed that, for the considered low to moderate filler contents, all the cases approximately follow a linear relationship with the CNT volume fraction. It is also noticeable that filler misalignment diminishes the fracture energy. Secondly, the effects of filler aspect ratio, filler content and CNT orientation distribution are quantified in Fig. \ref{Fig:WDfracture}. Fig. \ref{Fig:WDfracture}a shows the effects of the filler aspect ratio and the filler volume content upon the fracture energy of CNT-based composites, assuming random CNTs orientation. All curves increase until a particular aspect ratio is reached, $L_{cnt}/D_{cnt} \approx 370$ - the critical embedment length. Before and after this critical length the fracture behaviour is dominated by the CNT pull-out and the rupture mechanisms, respectively. Figure \ref{Fig:WDfracture}b shows the fracture energy as a function of the aspect ratio for different CNT mean angles $\theta_{\mu}$. It can be seen that the critical aspect ratio takes a maximum value when the CNTs are aligned with the load.

\subsubsection{Inhomogeneous CNT dispersions}

\begin{figure}[h]
\centering
\includegraphics[scale=1.0]{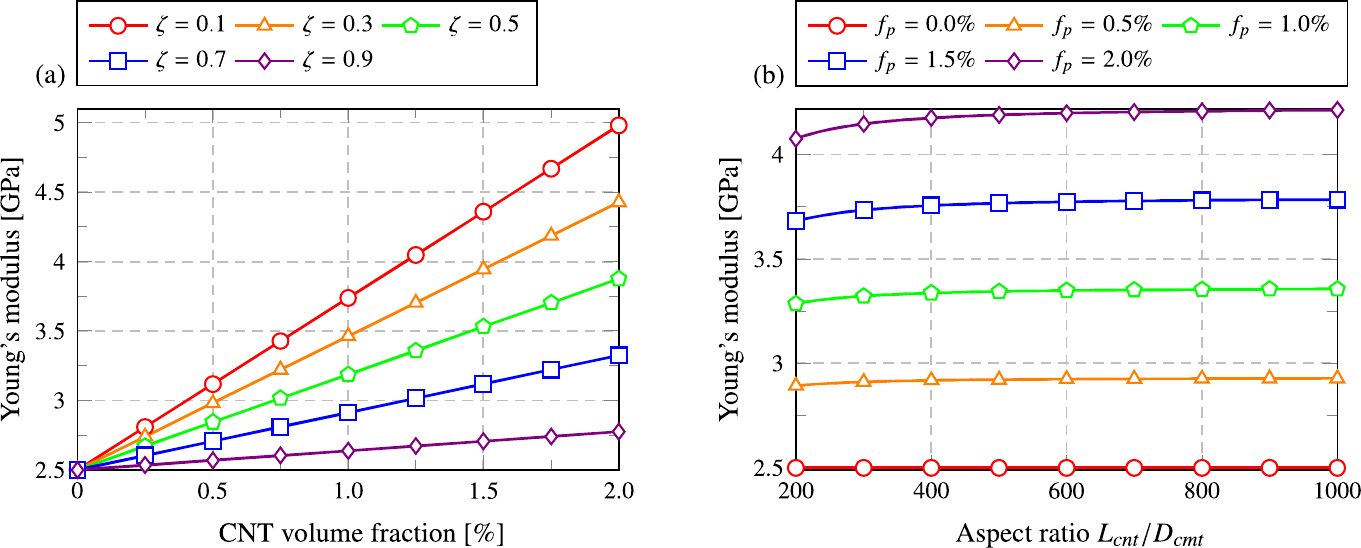}
\caption{Estimation of the effective elastic modulus of MWCNT/epoxy composites as a function of the volume fraction and considering agglomeration effects: (a) Effect of the agglomeration parameter $\zeta$ for a constant value of $\chi=0.2$ and $L_{cnt}/D_{cnt}=310$, and (b) effect of the filler aspect ratio $L_{cnt}/D_{cnt}$ considering agglomeration parameters $\zeta=0.4$ and $\chi=0.2$.}
		\label{Fig:two-step}
\end{figure}

In this section, we assess the theoretical approach proposed for the modelling of filler agglomeration effects upon the elastic moduli and fracture energy of CNT-based composites. The Young's modulus is shown in Fig. \ref{Fig:two-step}a as a function of the CNT filler volume fraction and the agglomeration parameter $\zeta$. The filler aspect ratio is approximately $L_{cnt}/D_{cnt}=310$ (see Table \ref{fittedparam}). As expected, all curves start at 2.5 GPa, the Young's modulus of the epoxy matrix. As the agglomeration parameter $\zeta$ increases, the curves exhibit a diminishing slope due to the agglomeration-induced loss in the effective stiffness of the composites. The selected agglomeration parameters range from $\zeta=0.2$, that is close to the case of uniformly dispersed CNTs, to $\zeta = 0.9$ where 90\% of CNTs are agglomerated forming bundles. The influence of the filler aspect ratio is shown in Fig. \ref{Fig:two-step}b, for different volume fractions and agglomeration parameters $\zeta=0.4$ and $\chi=0.2$. Qualitatively, the results of Fig. \ref{Fig:two-step}b resemble those presented in Fig. \ref{Fig:WDYoungAR}a, but significant quantitative differences are observed due to the effect of CNT agglomeration. The influence of the filler aspect ratio on Young's modulus is only evident for $L_{cnt}/D_{cnt}$ values between 200 and 400, after which the magnitude of Young's modulus tends to a constant value. 

\begin{figure}[h]
\centering
\includegraphics[scale=1.0]{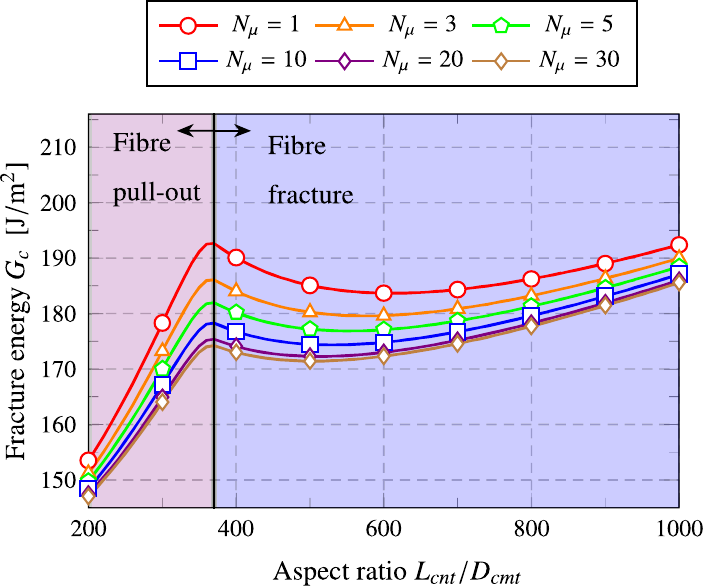}
\caption{Estimation of the fracture energy of MWCNT/epoxy composites as a function of the filler aspect ratio, considering different number $N_{\mu}$ of CNTs agglomerated in bundles.} 
\label{Fig:FE_aggl_diam}
\end{figure}

\begin{figure}[h]
\centering
\includegraphics[scale=1.0]{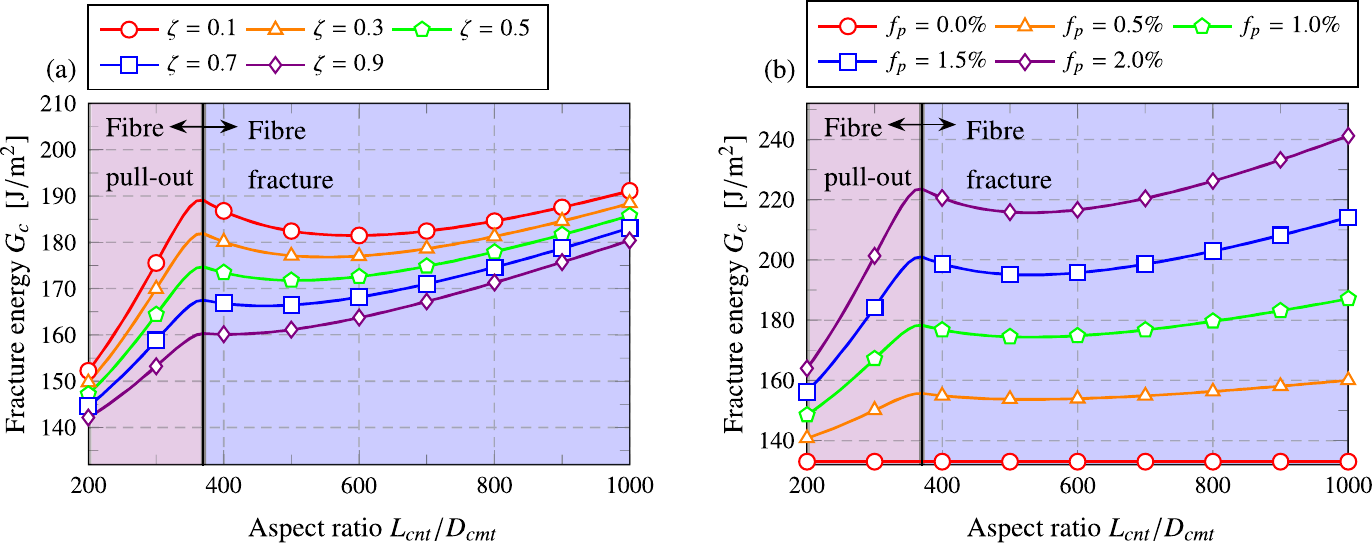}
\caption{Estimation of the fracture energy as a function of the aspect ratio in a MWCNT/epoxy composite, considering: (a) different values of the agglomeration parameter $\zeta$ ($f_{cnt}=1\%$), and (b) different filler volume fractions ($\zeta=0.4$ and $\chi=0.2$).} 
		\label{Fig:FE_aggl_zeta_f}
\end{figure}

The effect of filler agglomeration upon the fracture energy of CNT/epoxy composites is investigated in Figs. \ref{Fig:FE_aggl_diam} and \ref{Fig:FE_aggl_zeta_f}. The mean number of CNTs forming bundles ($N_{\mu}$) is a key factor, as shown in Fig. \ref{Fig:FE_aggl_diam}. Note that the case $N_{\mu}=1$ corresponds to the situation of an homogeneous dispersion of CNTs. Increasing $N_{\mu}$ values (i.e. more severe agglomeration) leads to substantial reductions in the effective fracture energy of composites when the filler aspect ratio falls within the range dominated by the pull-out mechanism ($\lessapprox 380$). This is due to the reduction in the filler/matrix interfacial area and, as a result, the decrease of the fracture energy contributed by fiber pull-out when CNTs form agglomerates - see Eq. \eqref{Eq:W_agg}. Nevertheless, such weakening effects decrease as the filler aspect ratio increases, and all curves come closer together for aspect ratios larger than about $L_{cnt}/D_{cmt}=800$. This is explained by the assumption that the formulation of the critical length $L_{c\theta}$ in Eq. (\ref{Lcrit}) remains valid in the case of agglomerated CNTs. When CNTs cluster together forming bundles, the diameter of the equivalent fibers increases and, as a result, so does the critical length $L_{c\theta}$. Although the fracture energy contributed by CNT fracture remains invariant and the energy by fiber pull-out decreases when CNTs form agglomerates, the increase in the critical embedment length slightly extends the number of fillers contributing through CNT pull-out after the critical embedment length. This fact, along with the reported stronger contribution of CNT pull-over over CNT fracture, explains the reduction in the detrimental effects induced by filler agglomeration observed in Figs. \ref{Fig:FE_aggl_diam} and \ref{Fig:FE_aggl_zeta_f}. This approach is deemed suitable for common CNT aspect ratios, which are typically below $L_{cnt}/D_{cmt}=1000$.

The effect of the agglomeration parameter $\zeta$ on the effective fracture energy of MWCNT/expoy composites is illustrated in Fig. \ref{Fig:FE_aggl_zeta_f}a for different filler aspect ratios. Notable reductions in the fracture energy of the composite are observed as the agglomeration parameter increases from $\zeta=0.1$ (uniform dispersion) to a highly agglomerated condition, $\zeta=0.9$. Finally, Fig. \ref{Fig:FE_aggl_zeta_f}b presents the fracture energy as a function of the filler aspect ratio for different filler volume fraction and agglomeration parameters $\zeta=0.4$ and $\chi=0.2$. The comparison with the results previously shown in the Fig. \ref{Fig:WDfracture}(a) for uniformly dispersed CNTs reveals agglomeration-induced reductions in fracture energy of about 30\%.

\subsection{Macroscopic crack propagation predictions}
\label{Section:5.1-Benchmarks}

\begin{figure}[h]
\centering
\includegraphics[scale=0.8]{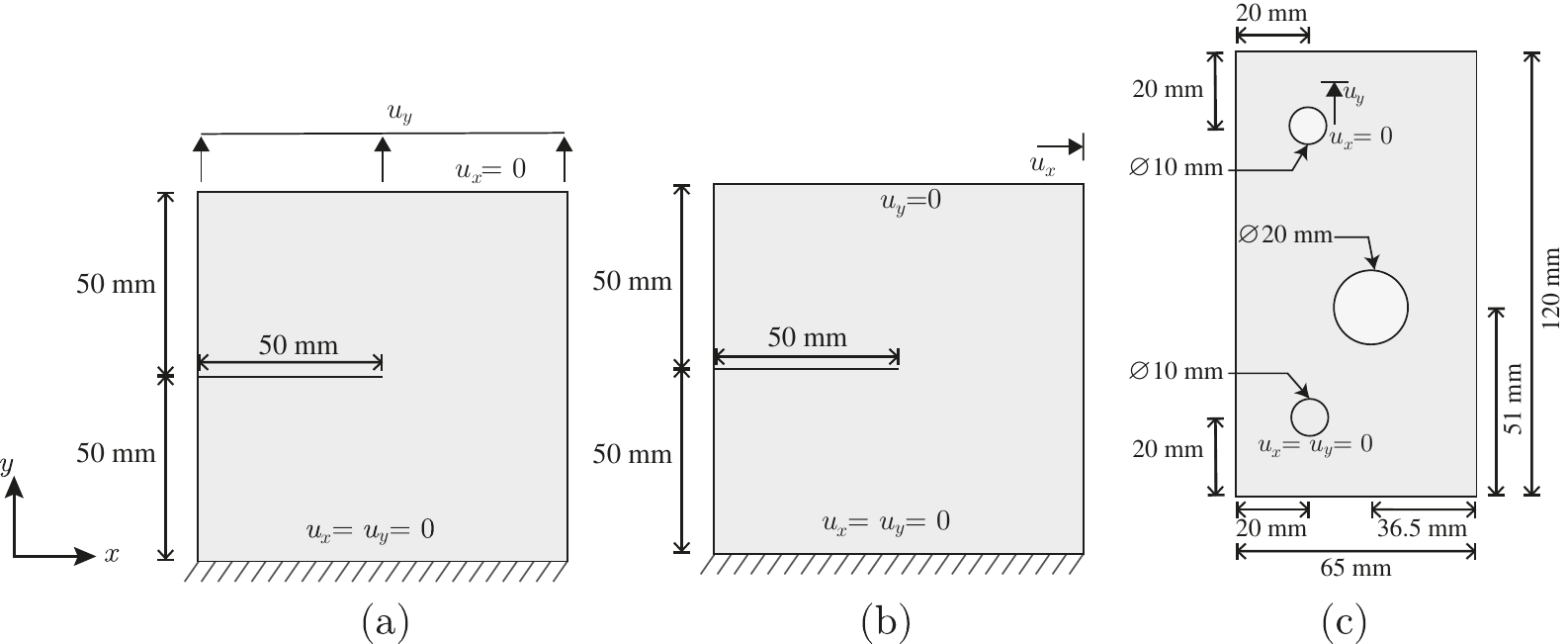}
\caption{Geometry and boundary conditions of the three case studies considered: (a) a notched plate in traction, (b) a notched plate under shear, and (c) a holed plate.}
\label{fig:test_illu}
\end{figure}

\begin{figure}[h]
\centering
\includegraphics[scale=0.8]{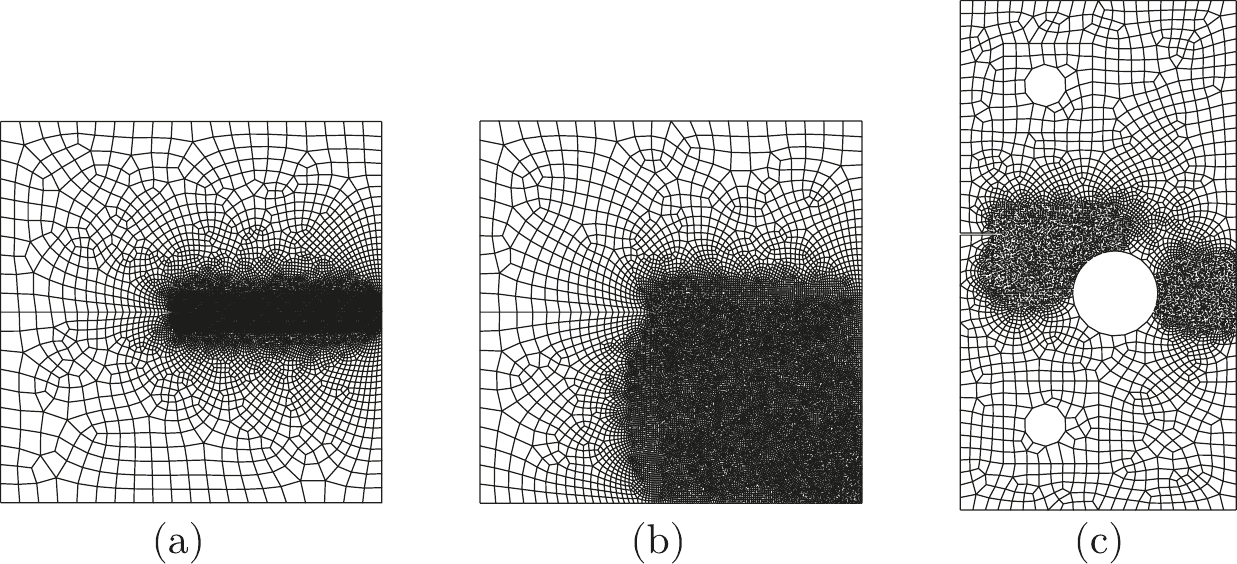}
\caption{Finite element mesh of the three case studies considered: (a) a notched plate in traction, (b) a notched plate under shear, and (c) a holed plate.}
\label{fig:test_illu2}
\end{figure}

We shall now demonstrate the potential of the proposed micromechanical phase field fracture approach for predicting the macroscopic fracture behaviour of CNT-based composites. Three case studies are considered: a single-edge notched plate subjected to uni-axial and shear loading, and a holed plate under traction. The geometrical configuration of these case studies is sketched in Fig. \ref{fig:test_illu}, and the material parameters used in the simulations are the ones reported in Table \ref{fittedparam}. The finite element meshes employed in each of the boundary value problems considered are shown in Fig. \ref{fig:test_illu2}; linear quadrilateral plane strain elements are used.  

\subsubsection{Single-edge notched specimen subjected to uniaxial tension}
\label{Section:5.1-axial-tension}

\begin{figure}[H]
\centering
\includegraphics[scale=1.0]{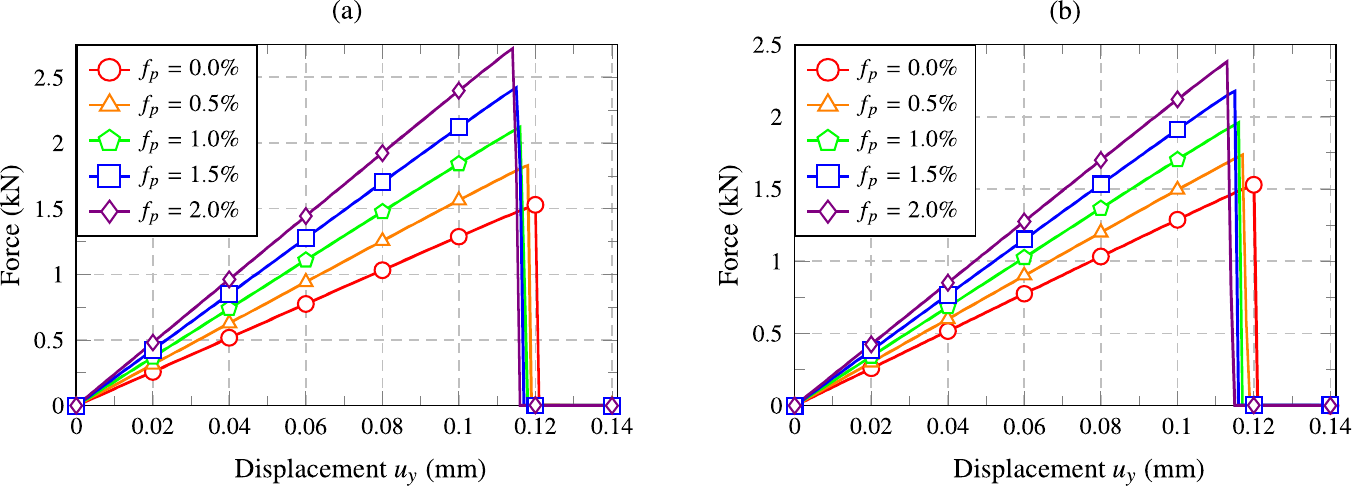}
\caption{Load-displacement curves of the notched specimen under traction for different CNT volume fractions, considering (a) uniform, and (b) inhomogeneous filler dispersions.} 
\label{Fig:5.1.test_1}
\end{figure}

\begin{figure}[H]
\centering
\includegraphics[scale=1.0]{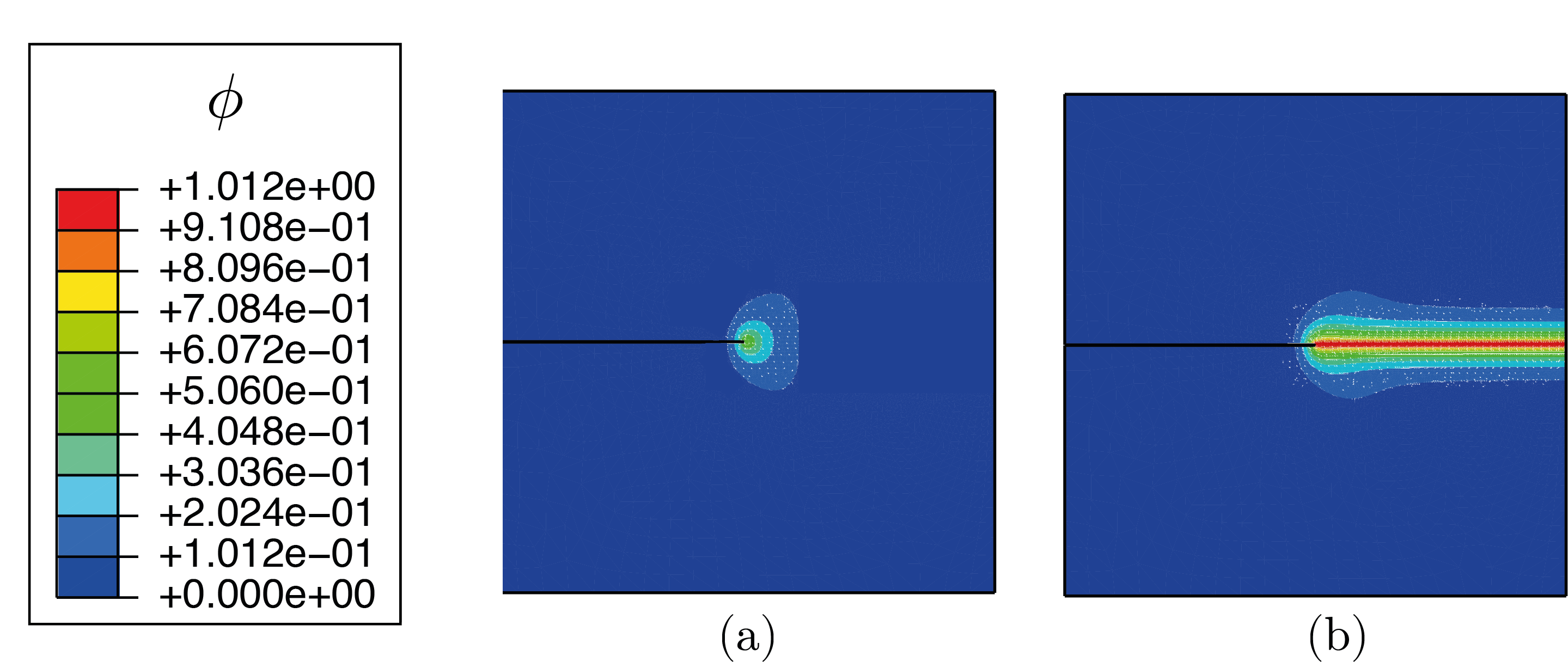}  
\caption{Phase field contours plots ($\phi$) of the notched plate under traction at remote displacement values of (a) $u_{y} = 0.1168$ $\rm{mm}$ and (b) $u_{y} = 0.117$ $\rm{mm}$. The results correspond to the case of uniformly dispersed CNTs with a volume fraction of 1\%.} 
\label{Fig:5.1.crack_traction}
\end{figure}

The paradigmatic benchmark of a square plate with a notch is addressed first. The geometry and finite element mesh are given in Figs. \ref{fig:test_illu}a and \ref{fig:test_illu2}a, respectively. The plate has an initial horizontal crack going from the left side to the centre of the specimen. The domain is discretised using a total of 8,532 elements, with the mesh refined along the expected crack propagation region. The characteristic element length is chosen to be 7 times smaller than the phase field length scale, to ensure mesh objectivity \cite{CMAME2018}. Here, $\ell=2.4$ mm. The load versus displacement curves obtained using homogeneous and inhomogeneous filler dispersions are given in Figs. \ref{Fig:5.1.test_1}a and \ref{Fig:5.1.test_1}b, respectively. In both cases, the force-displacement curve rises linearly until the critical load is reached, after which the failure process becomes unstable and the crack propagates across the plate in a sudden manner. Quantitative differences can be observed if agglomeration effects are accounted for - the maximum load for the plate doped with uniformly dispersed CNTs is 2.72 kN, which is $11.5\%$ higher than the one predicted in the case of inhomogeneous CNT dispersions. Regarding the influence of the CNT volume fraction $f_p$, a notable sensitivity is observed. The magnitude of the critical load increases significantly with $f_p$ due to fibre bridging toughening; e.g., the critical load can be up to 80\% higher for only a 2\% volume fraction, relative to the case of the epoxy matrix. The cracking pattern is qualitatively the same in all cases - a representative result for the case of a uniform dispersion of CNTs with a 1\% volume fraction is shown in Fig. \ref{Fig:5.1.crack_traction}.

\subsubsection{Single-edge notched specimen subjected to shear loading}
\label{Section:5.1-shear}

\begin{figure}[h]
\centering
\includegraphics[scale=0.9]{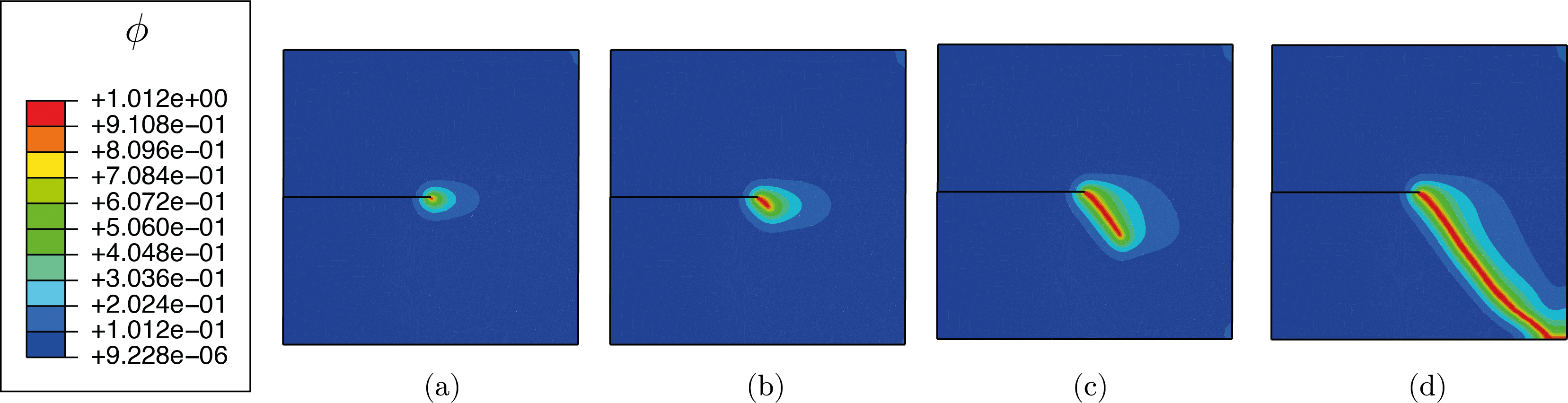}
\caption{Phase field $\phi$ contour plots corresponding to a value of the applied displacement equal to (a) $u_{x}=0.165$  $\rm{mm}$, (b) $u_{x}=0.195$ $\rm{mm}$, (c) $u_{x}=0.21$ $\rm{mm}$, and (d) $u_{x}=0.283$ $\rm{mm}$. The results correspond to the case of uniformly dispersed CNTs with a volume fraction of 1\%.}
\label{fig:crack_shear}
\end{figure}

\begin{figure}[h]
\centering
\includegraphics[scale=1.0]{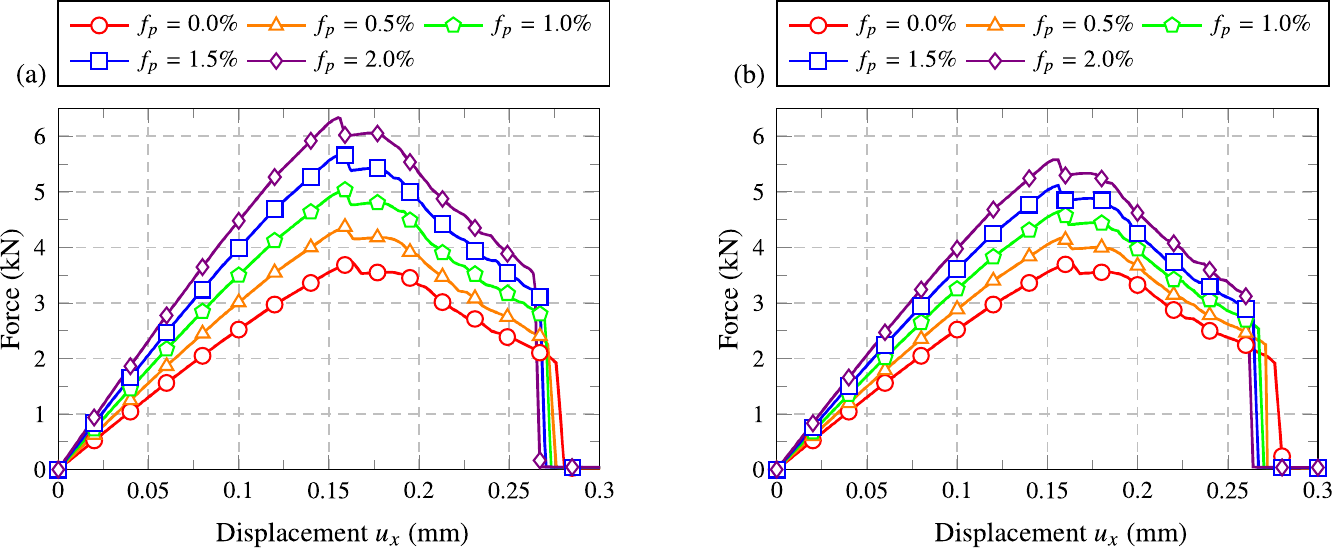}
\caption{Load-displacement curves of the notched specimen subjected to shear loading for different CNT volume fractions, considering (a) uniform, and (b) inhomogeneous filler dispersions.} 
\label{Fig:5.1.Notched_shear_vf}
\end{figure}

The same notched plate studied in the previous section is herein subjected to shear loading, as shown in Fig. \ref{fig:test_illu}b. The finite element mesh employed, shown in Fig. \ref{fig:test_illu2}b, uses 19,318 elements. The characteristic element length is 7 times smaller than the phase field length scale, which equals $\ell=2.4$ mm. Phase field contours for four stages of crack growth are shown in Fig. \ref{fig:crack_shear}. The load-displacement curves considering uniform and inhomogeneous CNT dispersions are shown in Figs. \ref{Fig:5.1.Notched_shear_vf}a and \ref{Fig:5.1.Notched_shear_vf}b, respectively. As in the previous case study, larger load carrying capacities are observed for CNT composites with higher volume fractions. In all the cases, the shear force applied to the plate reaches a maximum peak for an applied displacement of approximately $u_{x}=0.169$ $\rm{mm}$, when damage initiates near the crack tip. Afterwards, a diagonal crack propagates in a stable manner through the specimen towards the left bottom corner. Complete rupture and loss of load carrying capacity is observed at a remote displacement between 0.265 $\rm{mm}$ and 0.28 $\rm{mm}$, depending on the volume fraction and the filler dispersions. Note that CNTs agglomeration provokes noticeable reductions in the maximum forces sustained by the plate. For instance, a reduction of about 14.3\% is found in the case of the plate doped with 2\% CNT volume fraction.

\subsubsection{Holed plate under traction loading}
\label{PWH}

\begin{figure}[h]
\centering
\includegraphics[scale=1.0]{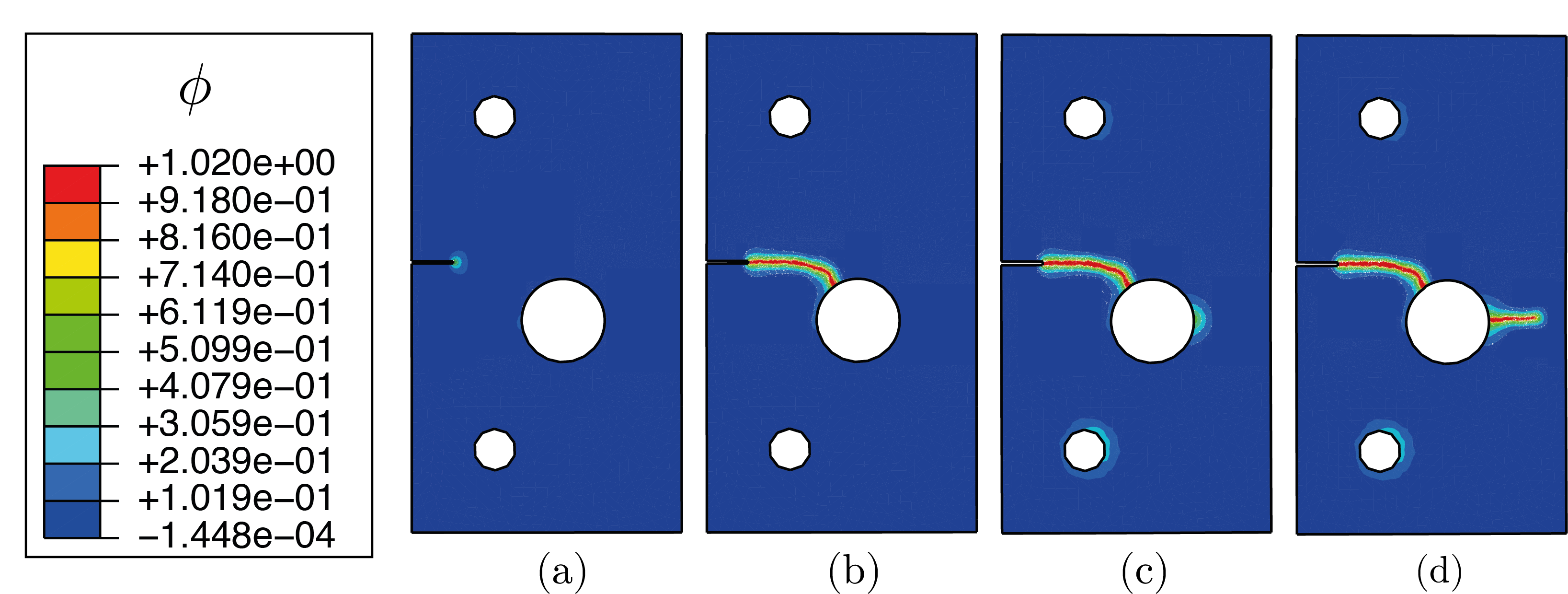}
\caption{Phase field $\phi$ contour plots in a holed plate under traction for different imposed displacements; namely (a) $u_{y}=0.104$ $\rm{mm}$, (b) $u_{y}=0.106$ $\rm{mm}$, (c) $u_{y}=0.299$ $\rm{mm}$, and (d) $u_{y}=0.320$ $\rm{mm}$. The results correspond to the case of uniformly dispersed CNTs with a volume fraction of 1\%.}
\label{fig:crack_platehole}
\end{figure}

Finally, we investigate the fracture response of a CNT composite holed plate with the geometry and boundary conditions shown in Fig. \ref{fig:test_illu}c. In this case study, the loading is applied through displacement controlled metal pins inserted into the two 10 $\rm{mm}$ diameter holes. The finite element mesh, shown in Fig. \ref{fig:test_illu2}c, contains 9,301 elements. The mesh is refined in the potential crack propagation regions to resolve the fracture process zone. Here, the phase field length scale equals $\ell=0.9$ mm. The cracking process is shown in Fig. \ref{fig:crack_platehole}, in terms of the phase field contours. Four stages are observed. First, the phase field increases its magnitude near the crack tip (Fig. \ref{fig:crack_platehole}a). Second, Fig. \ref{fig:crack_platehole}b, the crack propagates and deflects towards the hole. This is followed by stage three, Fig. \ref{fig:crack_platehole}c, when a second crack nucleates in the right edge of the hole. Finally, as shown in Fig. \ref{fig:crack_platehole}d, this second crack propagates until reaching the edge of the specimen. 

\begin{figure}[h]
\centering
\includegraphics[scale=1.0]{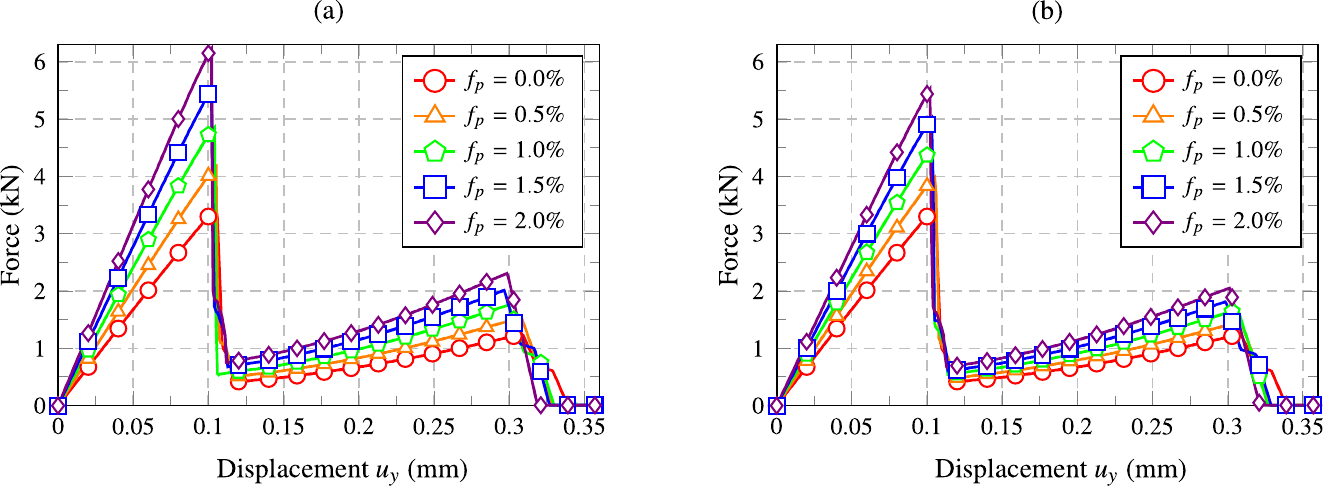}
\caption{Load-displacement curves of the holed plate under traction for different CNT volume fractions, considering (a) uniform, and (b) inhomogeneous filler dispersions.} 
\label{Fig:5.1.Notched_holer_vf}
\end{figure}

These four stages are clearly noticeable in the force versus displacement response, as shown in Fig. \ref{Fig:5.1.Notched_holer_vf}a for the case of uniform filler dispersions, and Fig. \ref{Fig:5.1.Notched_holer_vf}b for the case when aggregation effects are accounted for. The force increases until the first crack reaches the hole, when a sudden drop in the force versus displacement curve is observed. The displacement must increase significantly for the second crack to nucleate, but then its propagation is relatively fast and the plate loses its load carrying capacity completely. The mechanical response reveals a similar qualitative dependency on the micromechanics of the problem to that observed in the previous case studies. Higher CNT contents lead to higher stiffnesses and critical loads. For example, a CNT volume fraction of 2\% leads to a load bearing capacity of 6 \rm{kN}, almost twice the capacity of the plate made of pristine epoxy. Furthermore, the consideration of agglomeration effects translate into a reduction of the fracture resistance of the solid, with unstable fractures occurring at noticeably lower loads.


\section{Conclusions}
\label{Section: 6.conclusion}

We have presented a micromechanics- and phase field-based formulation for predicting crack nucleation and growth in carbon nanotube (CNT) composites. To achieve this, the model includes two key features: (i) a novel mean-field theory for determining the elastic and fracture properties of the composite, and (ii) a pioneering combination of analytical homogenisation and a phase field fracture formulation for CNT composites. Our homogenisation framework adopts a double-inclusion approach to into account filler/matrix interphase effects on the elastic properties and integrates the contribution of CNT pull-out and fracture mechanisms in the estimation of the composite's fracture resistance. Moreover, we also consider agglomeration effects on the elastic and fracture properties of CNT-based composites with a two-parameter agglomeration model. This includes the development of a new equivalent filler agglomeration approach for determining the macroscopic fracture properties. The developed model accounts for the separate contribution of isolated and clustered CNTs to the overall fracture energy. Clustered CNTs are simulated as equivalent fillers with cross-sections representing the agglomerated fillers, and the number of clustered fillers is treated in stochastic terms. The integration into a robust, monolithic phase field fracture framework enables predicting complex cracking phenomena in CNT-based composites and quantifying the role of CNTs on fracture resistance. The main findings of our analytical and numerical experiments are:
\begin{itemize}
    \item Filler agglomerates reduce the fracture resistance of CNT-based composites by facilitating CNT pull-out events. Composites doped with inhomogeneous dispersions of CNTs have lower filler/matrix interfacial areas and, as a result, sustain smaller pull-out forces.
    
    \item Increasing the CNT filler content raises the fracture resistance of CNT composites due to toughening through fibre pull-out and fibre fracture mechanisms. The critical load is found to increase by 80\% if a 2\% volume fraction of CNTs is incorporated.
    
    \item Accounting for the role of CNT fibre agglomeration is essential to quantitatively reproduce the toughness sensitivity to the CNT mass fraction reported in the experiments.
    
    \item The filler aspect ratio determines the dominant toughening mechanism. As a result, the fracture energy of the composite increases with filler aspect ratio until approaching the embedment length.
    
    \item While quantitative differences in terms of stiffness and fracture resistance are observed, crack trajectories and force versus displacement responses are found to be in good qualitative agreement.
\end{itemize}

The framework developed offers a pathway for designing fracture resistant CNT composite components undergoing cracking phenomena of arbitrary complexity. 

\section*{Acknowledgements}
\label{Sec:Acknowledgeoffunding}

L. Quinteros acknowledges financial support from the National Agency for Research and Development (ANID)/  Scholarship Program / DOCTORADO BECAS CHILE/2020 - 72210161. E. Mart\'{\i}nez-Pa\~neda was supported by an UKRI Future Leaders Fellowship (grant MR/V024124/1).







\bibliographystyle{elsarticle-num-nodoi}


\end{document}